\documentclass[aps,prb,twocolumn,showpacs,floatfix,superscriptaddress,amssymb,amsmath]{revtex4}
\usepackage{graphicx}
\usepackage{dcolumn}
\usepackage{bm}
\usepackage{psfrag}

\newcommand{\be}{\begin{equation}}
\newcommand{\ee}{\end{equation}}
\newcommand{\bea}{\begin{eqnarray}}
\newcommand{\eea}{\end{eqnarray}}

\newcommand{\s}{\sigma}
\newcommand{\vare}{\varepsilon}
\newcommand{\ri}{\mbox{i}}

\newcommand{\ba}{\begin{array}}
\newcommand{\ea}{\end{array}}
\def\nn{\nonumber\\}
\def\vare{{\varepsilon}}

\def\down{\downarrow}
\def\la{\langle}
\def\ra{\rangle}

\begin{document}

\title{Spin Hall effect in two-dimensional $p$-type semiconductors in  a
 magnetic field}

\author{Mahdi Zarea and Sergio E. Ulloa}
\affiliation{Department of Physics and Astronomy, and Nanoscale
and Quantum Phenomena Institute, \\Ohio University, Athens, Ohio
45701-2979}

\date{\today}

\begin{abstract}
We calculate the  spin Hall conductivity driven by Rashba spin-orbit 
interaction in $p$-type two-dimensional semiconductors
in the presence of a perpendicular magnetic field. For a highly confined
quantum well, the system is described by a $k$-cubic Rashba term for
 two-dimensional heavy holes. 
 The eigenstates of the system can be described by
 Landau spinor states. First we consider the conventional spin Hall 
conductivity. The contribution of the interband 
transitions  to the Kubo-Greenwood formula 
 gives the density dependent
 intrinsic  spin Hall conductivity, which approaches its  universal value
$\sigma^z_{xy}=9e/8\pi$ for weak spin-orbit coupling and low Fermi energies,
in agreement with previous work.
 However two intraband
contribution terms cancel this effect leading to zero conventional
spin Hall conductivity. Adding the torque dipole contribution to the definition
of spin current,
 we also study the {\it effective} spin conductivity. This
is shown to be proportional to the total magnetization 
plus surface terms which
exactly cancel it for small spin-orbit coupling.
If in low magnetic field the 
intraband transitions evolve to vertex corrections, the fact that both 
{effective} and conventional spin Hall conductivities vanish is unexpected. 
This suggests that
the zero magnetic field limit of the model is singular.    
\end{abstract}

 \keywords{}

\maketitle

\section{Introduction}

The goal of  spintronics is to inject and manipulate the spin of  
charge carriers in semiconductors. 
Spin Hall effect (SHE) has been proposed as an efficient source of
spin current in semiconductors. Two different mechanisms of
SHE are proposed.\cite{review}
In the extrinsic SHE the
skew scattering of electrons (driven by an electric field)
 from impurities with spin-orbit (SO)
 dependent scattering potential leads to opposite
spin accumulation in the transverse edges of the
sample. \cite{dya}
On the other hand, the intrinsic SHE  
arises from the SO term in 
the single-particle Hamiltonian of paramagnetic materials and
 is present even in the absence of impurities.  
Intrinsic SHE is predicted for two major systems: Murakami {\it et al.}
\cite{mur2} have predicted a dissipationless spin current in
 $p$-type semiconductors described by the
Luttinger model\cite{lut} of spin-3/2 holes. Sinova {\it et al.}\cite{Sinova}
have calculated spin Hall conductivity (SHC) for a 
two-dimensional  
electron gas with k-linear  Rashba\cite{Rashba} SO
coupling.

 Two seminal experiments have reported the observation of SHE. 
Kato {\it et al.}
 \cite{kato} have used Kerr rotation to observe spin accumulation in n-type 
GaAs.
 Wunderlich {\it et al.} \cite{wun} have observed the effect in  a clean 2D 
hole system
 using circularly polarized light. It is generally 
 believed\cite{eng} that in the former experiment the measured SHE is 
extrinsic while in
 the latter experiment (due to very low impurities concentration)
 it is an intrinsic effect.\cite{ber,mur,lei}

When impurities are
 present their contribution to intrinsic SHE is crucial. 
Extensive theoretical,\cite{ino1,ino2,dim,oleg,mish,schwab,kha2} 
 and numerical studies,\cite{sin2,sheng,nico} 
 show that for the model of 
2D electrons  with
 $k$-linear Rashba SO interaction,
 the SHE does not survive the 
thermodynamic limit if disorder induced
 vertex corrections are included. This is because
in this system the spin current is the time derivative of the magnetization
and vanishes when the magnetization saturates.\cite{dim,oleg} On the other
 hand for the Luttinger model of a hole  gas in which the SO coupling depends 
 nonlinearly in momentum, the vertex correction is calculated to be zero 
within the ladder approximation and for short-range scattering potential of
the impurities.
\cite{mur,ber,lei}
Adding Rashba SO interaction in  highly confined
quantum wells, makes the
 system be described by $k$-cubic Rashba SO terms.\cite{win2} 
The spin Hall conductivity of this system in the
clean limit approaches
 the universal  value of $\sigma=9e/8\pi$ for
weak SO coupling.\cite{sch} The vertex corrections of this model are
calculated,\cite{ber,lei,kha} and confirmed numerically,\cite{syn,sin2,che} to 
be zero.

 Rashba has used 
 an external magnetic field on a 2D 
electron system with $k$-linear Rashba SO term to show that the zero spin 
current
is an intrinsic property of the system and persists in the presence of 
impurities.\cite{Rashba2} 
This system is described by two series of Landau levels (spinors)
 separated by the SO dependent gap. The standard Kubo-Greenwood formula
has been used to calculate the SHC. It turns out that
the interband transitions
 contribute to the universal intrinsic SHE while including
the intraband transitions  cancels the effect. In the zero magnetic
field limit the latter contributions evolve into vertex corrections.
\cite{sin2}

The relation between the
 {\it conventional}
 definition of spin current (as the expectation value of the 
product of the spin operator and velocity) and spin accumulation is not
obvious. In systems with SO interaction the spin current is not conserved.
The spin density continuity equation contains the torque contribution which
 adds a torque dipole term to the conventional spin current 
term.\cite{cul,shi} 
The resulting {\it  effective} spin current  has been introduced as the 
proper definition
of the spin current. The corresponding { effective} spin Hall conductivity 
is shown to have opposite sign with respect to the conventional SHC for 
$k$-linear
and $k$-cubic Rashba models.\cite{shi}

Parallel to recent work,\cite{Rashba2,sch} we consider a 2D hole gas model
 in a perpendicular magnetic field in the absence of Zeeman coupling
 and find its exact eigenstates and eigenvalues. The Kubo-Greenwood 
formula is then used to calculate both the conventional and {effective}
 SHC of the model  for
different values of SO coupling
 and Fermi energies, including both interband and
intraband contributions.  We show that unexpectedly 
 both {effective} and conventional
SHC go to zero for
small value of SO coupling and then take a finite value as the SO coupling 
increases.
  Although calculations of  vertex
corrections for different types of scatterers have actually 
appeared recently, \cite{kha,shytov} more detailed comparison 
and analysis will be
desirable especially in the presence of  magnetic field.

\section{The Model}
In  2D hole systems, the Rashba spin-orbit splitting
(due to the structure-inversion asymmetry) of heavy hole states 
 is of third order in in-plane 
wave vector $k$ as compared to  Rashba spin splitting of electron states 
which is linear in $k$. \cite{win2} 
For a very confined quantum well
and not too high densities,  the Hamiltonian of heavy holes 
can be written as
\be
H_0=\frac{\hbar^2k_x^2}{2m}+\frac{\hbar^2k_y^2}{2m}
+\gamma_0\hbar^3(k_-^3\s^++k_+^3\s^-),\label{hole}
\ee
where $k_{\pm}=k_x\pm \ri k_y$, 
$\s^{\pm}=(\s_x\pm \ri \s_y)/2$.
The Pauli matrices $\s$ represent spin-$3/2$ heavy holes with 
effective mass $m$, and 
$\gamma_0$ is the Rashba SO coupling which without the lose of generality
we assume to be positive. \cite{symmetry} Out of two energy branches
\be 
\vare(k)=\hbar^2k^2/2m\pm\gamma_0\hbar^3k^3, \label{enb}
\ee 
the lower energy branch is bounded from below
only for
\be \gamma_0< 1/(3mk\hbar).\label{limit}
\ee 
For low Fermi energies (or low hole densities $n_D$), this condition is
satisfied and the SHC is calculated in the dc limit  neglecting the
impurity vertex corrections.\cite{sch} When the impurity scattering
energy $\hbar/\tau$ is comparable with the SO splitting energy 
 $\gamma_0(2m\vare_f)^{3/2}$, the SHC goes to zero. In the
opposite  limit of
a clean system 
$\hbar/\tau\ll\gamma_0(2m\vare_f)^{3/2}\ll\vare_f$,   
 the dc SHC  is
found to be dependent on $\gamma_0^2n_D$:
\be
\s^z_{xy}=\frac{9e}{16\pi m\gamma_0}(\frac{1}{k^+_f}-\frac{1}{k^-_f})\approx
\frac{9e}{8\pi}1/\sqrt{1-4\pi\hbar^2 n_Dm^2\gamma_0^2}.\label{fsh}
\ee
Here $k^+_f(k^-_f)$ is the Fermi momentum for the upper (lower) energy 
branch (\ref{enb}). 
Recent theoretical,\cite{ber,lei} and numerical,\cite{syn,sin2,che} 
work shows that
the vertex corrections to the SHC of the above model
are zero for short-range disorder interaction. This is in contrast to the 
$k$-linear
Rashba model in which the vertex corrections lead to complete cancellation of
the spin current.

\vskip .3cm

We consider the above homogeneous model in a 
homogeneous perpendicular magnetic field $\vec{ B}=-B\hat z$, 
in the absence of Zeeman coupling.\cite{mal} 
The kinetic momentum $ P=\vec{p}+\frac{e}{c}\vec{A}$ satisfies the commutation
relation $[P_y,P_x]=\ri m\hbar\omega_c$, where $\omega_c=eB/mc$. We 
use the symmetric gauge $A=-\frac{1}{2}\vec{r}\times\vec{B}$. The
Hamiltonian is
\bea 
H&=&\frac{P_x^2}{2m}+\frac{P_y^2}{2m}+\gamma_0(P_-^3\s^++P_+^3\s^-) \nn
&=&\hbar\omega_c\left((a^{\dag}a+\frac{1}{2})
+\gamma(a^3\s^++{a^{\dag}}^3\s^-)\right)\label{model}
\eea
in which the bosonic operators $a=P_-/\sqrt{2m\hbar\omega_c}$ and
$a^{\dag}=P_+/\sqrt{2m\hbar\omega_c}$
describe Landau levels. The 
dimensionless parameter $\gamma$ is defined 
by $\gamma=\gamma_0\sqrt{8m^3\hbar\omega_c}$.

 Similar to the $k$-linear Rashba
model in a perpendicular magnetic field,\cite{Winkler,Rashba2,egu,zu} 
the exact 
eigenvalues and eigenstates
of the above model can be found.
The SO term couples every state
$(\phi_{n},0)$  to the $(0,\phi_{n+3})$ leading to two new
$\lambda=\pm$ spinor states:
\bea
&&\psi^{\lambda}_n=\left(\ba{c}\cos\theta^{\lambda}_n \phi_{n-2} \\
\sin\theta_n^{\lambda}\phi_{n+1}\ea \right),\nn
&&E^{\lambda}_n/\hbar\omega_c=n+\delta/\cos2\theta_n^{\lambda},\label{states}
\eea
in which $\delta=-3/2$,
$\tan2\theta_n^+=\sqrt{n(n^2-1)}\gamma/\delta,
~\theta_n^-=\theta_n^+-\pi/2$ where $n\geq 2$. The states 
$\phi_{0\down},\phi_{1\down}$
and $\phi_{2\down}$ are not coupled to any other state, i.e for $n<2$
we have $\theta_n^+=\theta_n^-=0$.

\begin{figure}
\psfrag{E}{$E/(\hbar\omega_c)$}
\resizebox{.5\textwidth}{!}
{\rotatebox{0}{\includegraphics{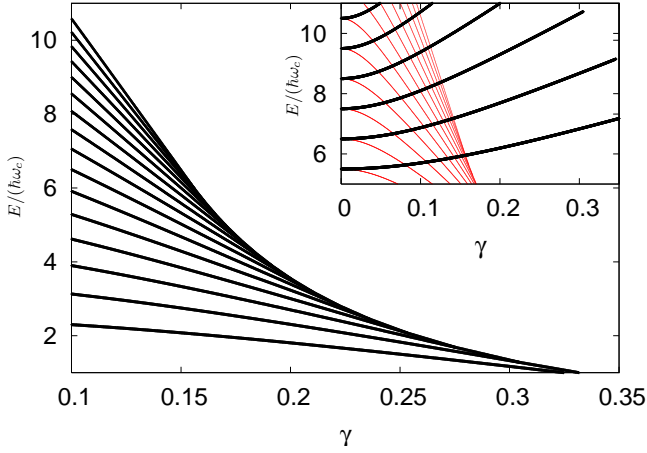}}}
\caption{ The energy levels $E^+_n$  of eq.(\ref{states})
as a function of $\gamma$ for $3<n<20$. Note that each energy
state $E^+_n$ is plotted as long as  $E^+_n>E^+_{n-1}$. 
The inset shows all states in that window, including
 $\lambda=-$ states that disperse upwards and so do not cross 
each other.}\label{level}
\end{figure}

We restrict the upper limit of SO coupling to $\gamma< 1/(3\sqrt{n_f})$ where 
$\vare_f=\hbar\omega_cn_f$ is the Fermi energy.  This condition is
 equivalent to equation (\ref{limit}). 
As long as this condition is satisfied, two 
$\lambda=+$ states do not cross each other. Energy levels are shown in
 Fig.\ \ref{level}; we can see that $\lambda=-$ states never cross
each other (for $\gamma>0$).

\section{The effective and conventional  spin current}

Using the commutation relation of $P_x$ and $P_y$ the velocity can be written 
as 
\bea
&&v_x=[x,H]/\ri\hbar=[P_y,H]/(\ri m\hbar\omega_c)\nn
&&v_y=[y,H]/\ri\hbar=-[P_x,H]/(\ri m\hbar\omega_c).\label{vel}
\eea
The same relation also holds for the linear SO system in magnetic field,
\cite{Rashba2} but the spin-dependent part of the velocity is
different in these two models. While in the hole system 
 velocity shows a $d$-wave symmetry,
$
v_x=P_x/m+3\gamma (P_xP_y+P_yP_x)\s^x+3\gamma(P_x^2-P_y^2)\s^y,
$
for the electron system we get $v_x=P_x/m+\gamma\s^y$ in which
the $\gamma$-dependent part is a constant. The symmetry of the 
velocity is directly reflected in the vertex corrections.  
For the 2D hole system in the absence of magnetic field this leads
to zero vertex corrections after integration over azimuthal angle
$\phi=\tan^{-1}(p_y/p_x)$. \cite{ber,lei}

The conventional definition of spin current for spin-$3/2$ holes
is
\be
J^z_x=\frac{3}{4}(v_x\s^z+\s^zv_x)=3P_x\s^z/2m. \label{spcu}
\ee
Note that in contrast to the $k$-linear Rashba model,
here the spin current operator can not be written as a commutator like
$J^z_x\sim [H,\s^x]$.\cite{dim} This is the profound reason behind
the vanishing of spin current in $k$-linear Rashba model for any
scalar random potential.

Relations (\ref{vel}) and (\ref{spcu}) show that 
$\la \lambda n|v_y|\lambda'n'\ra\la\lambda' n'|J_x^z|\lambda n\ra$ is
antisymmetric and so we can use  the following definition of conventional
SHC:\cite{Rashba2}
\be
\sigma^{z}_{xy}(\omega)=-3\ri\frac{em\omega_c}{\pi}\sum_{nn',\lambda\lambda'}^*
\frac{\la \lambda n|v_y|\lambda'n'\ra\la\lambda' n'|J_x^z|\lambda n\ra}
{(E_n^{\lambda}-E_{n'}^{\lambda'})^2-\hbar^2\omega^2}, \label{shc}
\ee
where the sum is restricted by 
$E_{n}^{\lambda}>\vare_f$ and  $E_{n'}^{\lambda'}<\vare_f$.
The Landau 
degeneracy factor  $m\omega_c/(2\pi\hbar)$ is also included.

Using the relation
$J^z_x=\frac{1}{\ri m\hbar\omega_c}[P_y\s^z,H_0]$ the conventional
spin
 current can be written as 
\be
J^z_x=\frac{-i}{2 m\hbar\omega_c}([P_y\s^z+\s^zP_y,H]
-[P_y\s^z+\s^zP_y,V])
\ee
where $V$ is the SO term in (\ref{model}). The first term in this formula is 
a full time derivative and
represents
 the {\it effective} spin current of the model (\ref{model}). By  simple
 algebra
the second
term can be written as $P_y\dot\s^z+\dot\s^zP_y$ which is the torque
dipole 
density. The dc SHC
is 
\bea
\sigma^{z}_{xy}(0)&=&\s^{z,s}_{xy}+\s^{z,\tau}_{xy}\nn
&=&\frac{\ri3e}{\pi m\omega_c\hbar^2}\sum_{nn',\lambda\lambda'}^*
\big(\la \lambda n|P_x|\lambda'n'\ra\la\lambda' n'|P_y\s^z|\lambda n\ra\nn
&-&\frac{\la \lambda n|P_x|\lambda'n'\ra
\la\lambda' n'|[P_y\s^z+\s^zP_y,~V]|\lambda n\ra}
{2(E_n^{\lambda}-E_{n'}^{\lambda'})}\big) \label{eshc}
\eea
in which the first term $\s^{z,s}_{xy}$ (respectively the second term 
$\s^{z,\tau}_{xy}$)  on the right hand side is the contribution of the 
{\it effective} spin current (respectively torque dipole density) 
to the conventional SHC.


\section{Results and discussion}

\begin{figure}
\resizebox{.5\textwidth}{!}
{\rotatebox{0}{\includegraphics{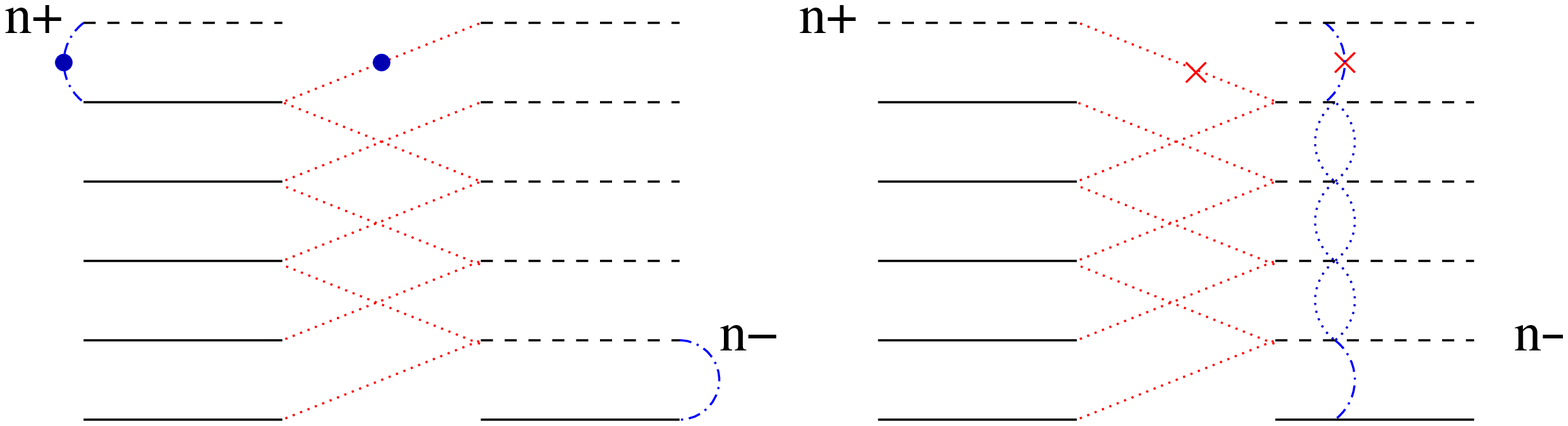}}}
\caption{(Color online)
 Left panel shows  possible  interband (dotted line) and 
intraband (dash-dotted line)
 transitions of the model. The solid (dashed) line shows filled (empty)
states. For convention the $\lambda=-1$ states are plotted lower.
 The right panel shows the equivalence of equations 
(\ref{my}) and (\ref{eshc}). Two subtracted $\times$ transitions and  
two added $\bullet$ transitions are~~`surface` terms in (\ref{my}).}\label{band}
\end{figure}

Relations (\ref{vel}) and (\ref{spcu}) indicate that the transition from the
state $n$ is allowed only to the state $n'=n\pm1$. 
In the energy spectrum (\ref{states}), the $\lambda=+$ state 
has lower energy than $\lambda=-$ levels (because we have defined $\delta<0$
and $\gamma>0$). 
For a given Fermi energy
 the first unoccupied $\lambda=+$ state $n^+$ is larger than the first
empty $\lambda=-$ state $n^-$ i.e always $n^+\geq n^-+3$.  
Therefore according to Fig.\ \ref{band} 
there are two contributions to the
spin Hall conductivity: the interband transitions occur from any state
 $\psi^-_n$
with $n\ge n^-$ to the state $\psi^+_{n\pm1}$ with $n\pm1<n^+$, 
and the intraband transition from 
$\psi^+_{n^+}$ to  $\psi^+_{n^+-1}$ and from $\psi^-_{n^-}$ 
to $\psi^-_{n^--1}$.

To calculate the  {effective} SHC in (\ref{eshc}) we apply 
the
same method used by Rashba,\cite{Rashba2} namely we write 
\bea
&&\sigma^{z,s}_{xy}(0)=3\ri\frac{e}{\pi m\omega_c\hbar^2}\nn
&&\big(\sum_{n=n^-}^{n^+-1}
\la -1,n|P_x|\lambda'n'\ra\la\lambda' n'|P_y\s^z|-1,n\ra\nn
&&+\la -1,n^+|P_x|+1,n^+-1\ra\la+1, n^+-1|P_y\s^z|-1,n^+\ra\nn
&&+\la +1,n^+|P_x|+1,n^+-1\ra\la+1, n^+-1|P_y\s^z|+1,n^+\ra\nn
&&-\la -1,n^+-1|P_x|-1,n^+\ra\la-1, n^+|P_y\s^z|-1,n^+\ra\nn
&&-\la -1,n^+-1|P_x|+1,n^+\ra\la1, n^+|P_y\s^z|-1,n^+-1\ra\big).\nn
\label{my}
\eea
In deriving this equation we have used the fact 
that the transition
$\psi^{-}_{n}\to\psi^{-}_{n+1}$ is canceled by
 $\psi^-_{n+1}\to\psi^{-}_{n}$.
 The equivalence of
(\ref{my}) and (\ref{eshc}) can be seen  from Fig.\ \ref{band}.
The sum of the last four `surface' terms in (\ref{my})
renders a constant $-3/2$ for the  $k$-cubic Rashba model using
(\ref{states}). On the other hand taking into account the $x-y$ symmetry 
of the model, the first term in (\ref{my})
can be written in the symmetric form 
$\sum_{n=n^-}^{n^+-1}
\la -1,n|[P_x,P_y]\s^z|-1,n\ra/2$. Finally we get the following 
relation for the { effective} SHC:
\bea
&&\sigma^{z,s}_{xy}(0)=\frac{3e}{\pi\hbar}\big(\sum_{n=n^-}^{n^+-1}
\la -1,n|\s^z|-1,n\ra/2+3/2\big)\nn
&&=3\frac{e}{\pi\hbar}(3/2-\sum_{n^-}^{n^+-1}\cos2\theta_n/2).\label{sh-ma}
\eea
The second term is  the
total
magnetization of the system in a magnetic field.

 For a given physical system the SO parameter $\gamma_0$ may depend on the 
intrinsic and applied electric fields.\cite{win2} The
Fermi energy depends linearly on the hole density $n_D$.
Moreover
$\gamma\sqrt{n_f}=(2m)^{3/2}\gamma_0\sqrt{\vare_f}$ is independent
of the applied magnetic field. In the accumulation layer of 
$GaAs-Al_{0.5}Ga_{0.5}As$ one obtains $\gamma\sqrt{n_f}\approx 0.3$ for
$n_D=10^{10}cm^{-2}$ while in a $200$\AA-wide  quantum well
 with
$n_D=10^{11}cm^{-2}$ we get $\gamma\sqrt{n_f}\approx 10^{-3}$.\cite{win2}

 We study the SHC as
function of this parameter. When plotted against this parameter 
all graphs of SHC for different Fermi energies (hole
densities)  will be scaled 
to one graph. To begin with,
the value of $\gamma$ is limited by $1<\gamma n_f^{3/2}<n_f$ which corresponds
to $\hbar/\tau_c<\gamma_0(2m\vare_f)^{3/2}<\vare_f$, i.e the same 
conditions as in the
free hole system in which the relaxation time $\tau$ is replaced by the 
cyclotron time $\tau_c=1/\omega_c$.  
For this value of $\gamma$ the mixing angle $\theta^\lambda$ approaches 
 its 
maximum value $-\pi/4$ and the conductivity (\ref{eshc})
 can be approximated as:
\bea
&&\sigma^{z,s}_{xy}(0)\approx-9\frac{e}{2\pi\hbar}
(1/\gamma\sqrt{n^+}-1/\gamma\sqrt{n^-}+1).\label{aeshc}
\eea

\begin{figure}
\resizebox{.5\textwidth}{!}
{\rotatebox{0}{\includegraphics{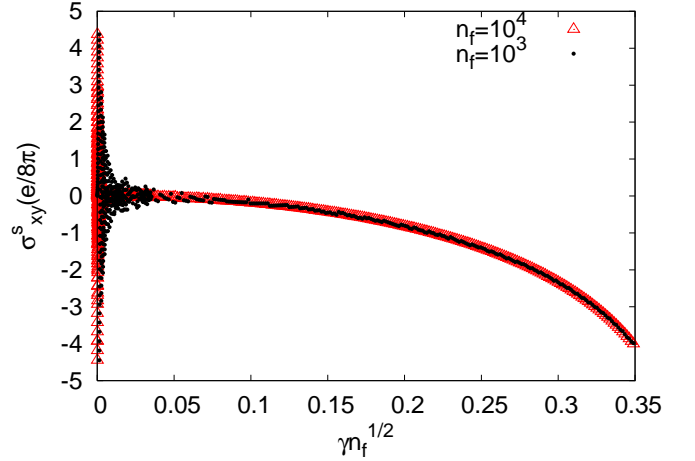}}}
\caption{(Color online)
 The {\it effective} spin Hall conductivity as a function of 
$\gamma\sqrt{n_f}=\gamma_0\sqrt{\vare_f}$ for different values of 
Fermi energy. All graphs scale to one if plotted against 
$\gamma\sqrt{n_f}$. For weak SO coupling or low Fermi energies the conductivity
approaches zero but it shows Shubnikov-de Haas oscillations.}\label{eshc1}
\end{figure}

In Fig.\ \ref{eshc1} we have plotted the {effective} SHC 
(\ref{eshc})
as a function of $\gamma\sqrt{n_f}$. For larger value of  $\gamma\sqrt{n_f}$
 the {effective} SHC $\sigma^{z,s}_{xy}$ is {\it negative} and {\it finite}. 
The
negative sign of the {effective} SHC for larger $\gamma\sqrt{n_f}$ is the
result of the original choice for positive conventional SHC.\cite{Rashba3}

\begin{figure}
\resizebox{.5\textwidth}{!}
{\rotatebox{0}{\includegraphics{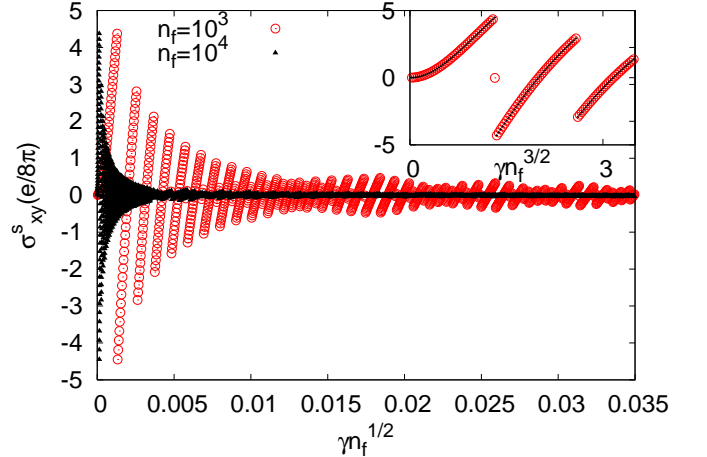}}}
\caption{ (Color online)
The {\it effective} spin Hall conductivity for very low value of 
$\gamma\sqrt{n_f}\sim\gamma_0\sqrt{\vare_f}$. The higher the $n_f$ values
 the smaller is the amplitude of oscillations, as one would expect.
The inset is exactly the same graph when plotted against 
$\gamma{n_f}^{3/2}$ which shows that the highest jump in SHC is
$\Delta\s^s_{xy}=9e/(8\pi)$ and it happens at the same value of
$\gamma{n_f}^{3/2}$ for any $n_f$.}
\label{eshc0}
\end{figure}

For lower values of $\gamma\sqrt{n_f}$ both the { effective} 
and conventional
 SHC approach zero. In Fig.\ \ref{eshc0} we have plotted the { effective} 
SHC for
 lower values of $\gamma\sqrt{n_f}$. Going to the zero limit the 
typical 
Shubnikov-de Haas 
oscillations appear in the conductivity. Notice that for
 smaller $n_f$ 
the amplitude of the oscillations is larger, as one would expect for
the quantum limit. 
 
\begin{figure}
\resizebox{.5\textwidth}{!}
 {\rotatebox{0}{\includegraphics{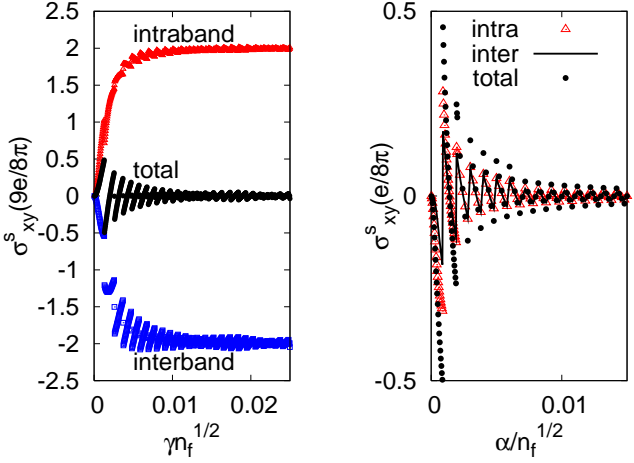}}}
\caption{ (Color online)
The distinct
 intra and interband transition contributions to the 
{\it effective} SHC for very low value of 
$\gamma\sqrt{n_f}$ in the $k$-cubic model (left panel). The right panel
 is the same for the
  $k$-linear 
model in which two contributions are zero independently.    }\label{comp}
\end{figure}

It is important to note that  the same
behavior can be observed for {effective} SHC in the linear Rashba
model in a magnetic field. 
Actually the formula (\ref{eshc}) is valid
for $k$-linear Rashba model (up to the factor of 9 which comes from spin-3/2 
holes and  $k^3$ momentum dependence). 
In Fig.\ \ref{comp} we  compare
the {effective} SHC of these two models. There are two differences between
the two systems. Firstly
 in the $k$-linear Rashba model with $V_{SO}=\alpha(k\times\s)$ 
the {\it effective} SHC remains  zero for
larger value of SO coupling $\alpha$ 
while in the $k$-cubic model it goes to finite negative
values (see Fig.\ \ref{eshc1}). 
However the deviation from zero in the latter case
 is because we  approach
the limit of unbounded negative energies while for the linear model the typical
value of SO coupling is very far from the upper limit $\alpha>\sqrt{n_f}$ of
 the unphysical region.
Second and more importantly, in the linear model both the interband and
intraband contributions to the {effective} SHC are zero {\it independently}
 while for the
$k$-cubic model these two contributions are finite but opposite in sign and 
cancel each other out. 

\begin{figure}
\resizebox{.5\textwidth}{!}
{\rotatebox{0}{\includegraphics{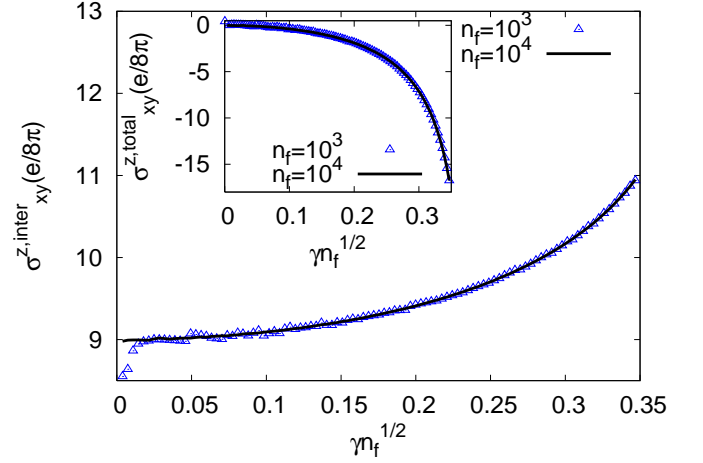}}}
\caption{ (Color online)
The interband contribution to 
the  {\it conventional} SHC as a function of 
$\gamma\sqrt{n_f}=\gamma_0\sqrt{\vare_f}$ for the $k$-cubic model. 
For weak SO coupling or low Fermi energies the conductivity
approaches its universal value $\s^z_{xy}=9e/8\pi$. When the 
intraband contribution is included,
the total conventional SHC is zero for small  $\gamma\sqrt{n_f}$ and goes to
negative values as  $\gamma\sqrt{n_f}$ increases. }\label{intf}
\end{figure}

For comparison we also calculate the {\it conventional} SHC (\ref{shc}).     
A straightforward 
replacement of the wave functions and energies in  (\ref{shc})
renders
\bea
&&\sigma^z_{xy}(0)=\frac{3e}{4\pi\hbar}\sum_{n,\lambda\lambda'}^*\nn
&&
\frac{  (n-1)\cos^2\theta_{n+1}^{\lambda'}\cos^2\theta_{n}^{\lambda}
  -(n+2)\sin^2\theta_{n+1}^{\lambda'}\sin^2\theta_{n}^{\lambda}}
{1+\delta/\cos2\theta_{n+1}^{\lambda'}-\delta/\cos2\theta_{n}^{\lambda}}\nn
&&-
\frac{  (n-2)\cos^2\theta_{n-1}^{\lambda'}\cos^2\theta_{n}^{\lambda}
  -(n+1)\sin^2\theta_{n-1}^{\lambda'}\sin^2\theta_{n}^{\lambda}}
{1+\delta/\cos2\theta_n^{\lambda}-\delta/\cos2\theta_{n-1}^{\lambda'}}\nn
\label{cur}
\eea
in which the first (respectively second) term describes the transition
 $n\to n+1$ (respectively $n\to n-1$).
 We first include only the  interband contribution to the conventional 
SHE to compare it to the system of free holes in the absence of magnetic field.
The result is plotted in Fig.\ \ref{intf} and it is actually the same as the
SHC for a free 2D electron gas.\cite{sch} For the range of $\gamma\sqrt{n_f}$
chosen here, (\ref{cur}) can be approximated by 
\bea
\sigma^{z,inter}_{xy}(0)&\approx& \frac{3e}{16\pi\hbar}\int_{n^-}^{n^+}
\frac{-3}{\gamma\sqrt{n^3}}dn\nn
&=& \frac{9e}{8\pi\hbar\gamma}
(\frac{1}{\sqrt{n^+}}-\frac{1}{\sqrt{n^-}}).\label{int}
\eea
which  using  $\hbar\omega_c n^{\pm}=\hbar^2k_{\pm}^2/2m$ 
 renders exactly
  (\ref{fsh}). 
Here, as $\gamma$ decreases the {\it conventional} SHC 
approaches its
  universal value $\sigma^z_{xy}=9e/8\pi$.\cite{sch}
Now we include also  the intrabandr transitions, consisting of 
the two terms
near the Fermi surface.
Their contribution is negative and cancels the conventional SHC for the lower
values of $\gamma\sqrt{n_f}$, as it is shown in the inset of Fig.\ \ref{intf}

\vskip .5cm
\section{Conclusion} 

We have studied the 2D hole gas system with $k$-cubic Rashba SO
interaction in a perpendicular magnetic field in the absence of
Zeeman splitting. The exact eigenstates and eigenvalues are derived and used
in a 
 Kubo formula to calculate  the { effective} and conventional SHC of 
the model. 
For the {\it conventional} SHC the interband transitions reproduce the
intrinsic spin Hall conductivity which approaches the universal value
of $\sigma^z_{xy}=9e/8\pi$ for weak SO coupling.\cite{sch} However 
when the negative intraband
contributions are included the {\it conventional} SHC becomes negative and
goes to zero for weak SO coupling. 

We also consider the {\it effective}
spin current in which the torque dipole is included. We show that 
 the  corresponding {\it effective} SHC shows Shubnikov-de Haas 
oscillations around zero for
small SO coupling, if both interband and intraband contributions are 
included.
Higher $n_f$ results in smaller oscillation amplitude. For stronger
SO coupling the {\it effective} SHC becomes negative.

Comparing the 
$k$-cubic model with the  $k$-linear model, the main difference
is that in the former the intraband and interband contributions to the
{\it effective} SHC are finite but opposite in sign and cancel each other.
 In the latter 
each contribution is zero independently.

 The fact that for low values of SO coupling
both the { effective} and conventional SHC go to zero is unexpected.
For the $k$-linear model it is well-known that the intraband transitions
evolve into the vertex correction which is proved to be zero.
For the $k$-cubic model, 
although it is
well known\cite{ber,lei,mur} that the vertex corrections are zero 
 in the case of short-range s-wave scatterers,
the relation between 
 the intraband transitions and the vertex corrections is not straightforward.
For the {\it effective} SHC  eq.(\ref{sh-ma}) suggests that  
the zero magnetic field limit is  singular in this model.
However more detailed calculations of the vertex and higher order 
corrections may help
to ascertain this behavior.

The above results would be modified  for an {inhomogeneous} system and
for the ac SHC $\s^z_{xy}(\omega)$, because the intraband
contributions are less stable and are reduced by frequency and
scattering.\cite{mish}
One expects that the intraband contributions will diverge 
at $\omega=\omega_c$ after which they will vanish for
$\omega_c\ll\omega$.

{\bf Acknowledgment}

 The work was supported by CMSS at Ohio University and NSF-NIRT.


\end{document}